\documentclass[epj]{svjour}
\usepackage{latexsym}
\usepackage{graphics}
\usepackage{epsfig}
\begin{document}
\title{Antiproton-deuteron annihilation at low energies}
\author{V.A. Karmanov\inst{1}, K.V. Protasov\inst{2} \and A.Yu. Voronin\inst{1}% etc
}                     % Do not remove

\institute{Lebedev Physical Institute,
Leninsky prospekt, 53, 117924 Moscow, Russia
\and
Institut des Sciences Nucl\'eaires, IN2P3-CNRS, UJFG,
53, Avenue des Martyrs, F-38026 Grenoble Cedex, France
}
\date{Received: date / Revised version: date}

\abstract{
Recent experimental studies of the antiproton-deuteron system at low 
energies have shown that the imaginary part of the antiproton-deuteron 
scattering length is smaller than the antiproton-proton one.  Two- and 
three-body systems with strong annihilation are investigated and a 
mechanism explaining this unexpected relation between the imaginary 
parts of the scattering lengths is proposed.
\PACS{
      {25.43.+t}{Antiproton-induced reactions}
     } } %end of abstract
\maketitle

\section{Introduction}

\label{intro} Recent measurements of 
antiproton-deute\-rium and anti\-proton-helium annihilation cross-section 
at low antiproton momentum ($p_{\mbox{\tiny lab}}$ down to 35 MeV/$c$) 
\cite {Zenoni99} and of the shift and the width of the 
antiproton-deuteron atomic ground state \cite{Augsburger99} have provided the 
first information on interaction of slow antiprotons with light nuclei. 
Both experiments gave unexpected but coherent results. The width of the 
1S state of the \={p}D atom appears to be approximately equal to that 
for protonium. In the region of low antiproton momentum, the 
annihilation cross-sections in the \={p}p, \={p}D and \={p}$^{4}$He 
systems are approximately equal.

The phenomenological analysis of these data \cite{PBLZ} allows 
the imaginary parts of the S-wave scattering lengths for the three 
nuclei to be extracted. The values was found to decrease 
with atomic number 
\begin{eqnarray*}
\mbox {Im } a_{sc} (\bar{\mbox{p}}\mbox{p})&=& - [0.69 \pm 0.01 (\mbox{stat}
) \pm 0.03 (\mbox{sys})] \mbox { fm}, \\
\mbox {Im } a_{sc} (\bar{\mbox{p}}\mbox{D})&=& -[0.62\pm0.02(\mbox{stat}
)\pm0.05 (\mbox{sys)}] \mbox{ fm}, \\
\mbox {Im } a_{sc} (\bar{\mbox{p}}^4\mbox{He})&=& -[0.36\pm0.03(\mbox{stat})
^{+0.19} _{-0.11}(\mbox{sys})] \mbox { fm}.
\end{eqnarray*}
This result is in direct contradiction with the naive geometrical picture 
of annihilation which would suggest that the imaginary part of the 
scattering length (or corresponding width and cross-section) increases 
with the size of the nucleus.

Only a few calculations of the atomic \={p}D system 
\cite {Wycech85,LT,Richard} have been carried out prior to the
measurements. In the first one \cite{Wycech85}, the authors used an 
approximate iterative method to obtain the optical \={p}D potential 
starting from elementary \={N}N and NN interactions. This method was 
applied previously to atomic data for heavier nuclei  \cite{WG,WGS}. In 
the second calculation \cite{LT}, the \={p}D system interaction was studied 
by solution of the three body equations. In both approaches, the 
elementary \={N}N and NN interactions was chosen in separable form 
(this limits the conclusions due to particular properties of 
the separable interaction). In the two articles, the same numerical 
result was obtained: the width of the 1S \={p}D atomic level appeared 
to be smaller than that for the \={p}p system and the necessity to push the 
calculations beyond the simple impulse approximation was emphasized. The 
same conclusion was obtained in \cite{Zen99}. In a third the article 
 \cite{Richard}, the authors have investigated mostly different 
approximate solutions of the three-body problem. Due to the strong 
annihilation the results was found to be not very sensitive to the 
three-body dynamics.  However, the shifts and the widths of the 1S 
\={p}D atomic level obtained in this article are quite different from 
the ones observed experimentally.

The aim of the present work is to understand why the simple geometrical 
picture does not work and to propose a mechanism, at least for \={p}D 
system, to explain the unexpected behavior of the scattering length.

Our explanation is based on two points. The first one 
is the fact that the imaginary part of the scattering length for the 
\={N}N optical potentials with strong annihilation is mainly determined 
by the diffuseness of the exponential tail of the potential, but not by 
the details of inner part of interaction. Therefore one can expect that 
$\mbox {Im }a_{sc}(\bar{\mbox{p}}\mbox{p})$ and $\mbox {Im 
}a_{sc}(\bar{\mbox{p}}\mbox{D})$ can be, at least, close to each other, 
in spite of the different structure of interactions in \={p}p and in 
\={p}D systems. This expectation is confirmed by our three-body 
calculation of $\mbox {Im }a_{sc}(\bar{\mbox{p}}\mbox{D})$.
The second one is the increase of $\mbox {Im }a_{sc}(\bar{\mbox{p}}\mbox{p})$
relative to $\mbox {Im }a_{sc}(\bar{\mbox{p}}\mbox{D})$ due to
a resonance in \={N}N system.

We shall be concerned with general features of annihilation in the \={p}p and 
\={p}D systems which are essential for explaining the observed phenomenon.
There are two 
possible methods for taking the annihilation into account:  either to 
introduce the imaginary part to the potential (optical model) or to 
introduce explicitly effective annihilation channels (coupled channel 
model). The first possibility is more simple but leads to some 
problems related to the non unitarity of this approach (see \cite{CDPS} 
and references therein). The second takes the unitarity into 
account explicitly but the numerical treatment becomes quite involved.  Thus,
to 
simplify the calculations only the  
optical model approach is considered. Furthermore, a very simple
antinucleon-nucleon 
interaction is used. 

To simplify the numerical treatment we will calculate the scattering
length $a_{s}$ corresponding to a pure strong interaction.
The scattering lengths extracted from the experimental data $a_{sc}$
are modified by the Coulomb interaction. 
For the \={p}p system, there is a simple phenomenological relation
between $a_{s}$, $a_{sc}$, and the Bohr radius $B$:
\begin{eqnarray*}
\frac{B}{a_{sc}}= \frac{B}{a_{s}} + C 
\end{eqnarray*}
with complex constant $C=7.17 + i1.46$ which is valid within 1\% accuracy as 
was demonstrated in \cite{KP}. If one supposes that this relation is 
the same for the \={p}D system one obtains the following imaginary 
parts of the scattering lengths:
\begin{eqnarray*}
\mbox {Im } a_{s} (\bar{\mbox{p}}\mbox{p})&=& -(0.86 \pm 0.04) \mbox { fm}, \\
\mbox {Im } a_{s} (\bar{\mbox{p}}\mbox{D})&=& -(0.80\pm0.06) \mbox{ fm}.
\end{eqnarray*}
The real parts of the scattering lengths are taken from \cite{PBLZ}, 
all errors are added quadratically. One can see that in absence of Coulomb 
forces the \={p}D scattering length is still smaller than that for the 
\={p}p system. 

The article is organized as follows. Section \ref{sec:1} presents the 
properties of simple two-body interactions in the case of strong 
annihilation. In Section \ref{sec:3}, the three-body \={p}D problem is 
solved numerically by means of Faddeev equations and a mechanism 
explaining the observed effect is found. Finally, we provide 
a brief summary of the results.

%%%%%%%%%%%%%%%%%%%%%%%%%%%%%%%%%%%%%%%%%%%%%%%%%%%%%%%

\section{Two-body interactions}
\label{sec:1}
Before calculating the \={p}D scattering length obtained in the next 
section by the numerical solution of the three-body problem it is 
instructive to present some analytical results on the scattering length 
in a few simple two-body optical potentials. These examples emphasize 
that the diffuseness of the interaction determines the imaginary 
part of the scattering length. This remains valid for \={p}D 
interaction.

For the square well potential 
\begin{eqnarray*}
U(r)=\left\{ 
\begin{array}{cl}
-U_{0}, & r<R \\ 
0, & r>R
\end{array}
\right. 
\end{eqnarray*}
 the scattering length is equal to
\begin{eqnarray}
\label{slsquare}
a_{s} =R\left( 1-\frac{\tan pR}{pR}\right) . 
\end{eqnarray}
Here $p^{2}=2mU_{0}$ with $U_{0}\equiv W\mbox{e}^{i\varphi }$. In the 
case of pure imaginary optical potential, $\varphi =\pi /2$. Hereafter, 
we suppose that $\varphi >0$. Potentials with an attractive real part have 
$0<\varphi <\pi/2$, and repulsive potentials have $\pi /2<\varphi <\pi $.
For a pure 
imaginary potential, the imaginary part of the scattering length as a 
function of the imaginary depth starts from zero, reaches a maximum 
and, in the limit of very strong annihilation ($W\rightarrow \infty $), 
tends to zero as $1/\sqrt{W}$. The scattering length becomes real.  
This result is well-known in physics of hadronic atoms \cite{Batty83}.  
The scattering on this potential is equivalent to scattering on hard 
sphere.  In this limit, the wave function tends to zero within the region 
of annihilation. This is the so-called ``S-wave suppression'' which appears 
in the optical potential with strong annihilation.

Even in the limit when the range of interaction becomes very large 
($R\rightarrow \infty $), the imaginary part of the scattering length 
does not go to infinity: 
\begin{eqnarray*}
\lim_{R\rightarrow \infty }\mbox{Im }a_{s}=-\frac{1}{\sqrt{2mW}}\,
\cos \frac{\varphi }{2}. 
\end{eqnarray*}
The parameter which defines the strength of annihilation is a product 
$\left| pR\right|$ (see (\ref{slsquare})).  So hereafter, the strong 
annihilation limit should be taken as implying $\left| pR\right| \gg 1.$

The fact that the imaginary part of the scattering length becomes real 
when the annihilation is strong is related to the sharp edge of the 
square well potential. For another analytically solvable potential 
\begin{eqnarray*}
U(r)=-U_{0}\mbox{e}^{-r/r_{0}} 
\end{eqnarray*}
with $U_{0}\equiv W\mbox{e}^{i\varphi }$, the scattering length
is equal to 
\begin{eqnarray*}
a_{s} = 2r_0 \left[ \gamma + \ln z_0 - \frac{\pi}{2}
\frac{N_0(2z_0)}{J_0(2z_0)} \right].
\end{eqnarray*}
Here $\gamma = 0.577...$ is Euler constant, 
$z_{0}^{2}=2mU_{0}r_{0}^{2}$, $J_0(z)$ and $N_0(z)$ are Bessel and 
Neumann functions of index 0.

In the limit of strong annihilation ($\left| z_{0}\right| \gg 1$), 
$\mbox{Im }a_{s}$ is defined by the diffuseness of the interaction 
$r_{0}$:
\begin{eqnarray*}
\lim_{\left| z_{0}\right| \rightarrow \infty }\mbox{Im }a_{s}=-r_{0}(\pi
-\varphi ). 
\end{eqnarray*}

The same asymptotic result for $\mbox{Im }a_{s}$
is valid for any complex potential which has an ``exponential tail'',
for instance, a Woods-Saxon potential which can be considered as a first
approximation to describe the antipro\-ton-nucleus atomic data
\cite{Batty8190}. More sophisticated potentials containing more
complicated dependencies are also discussed
in the literature (see 
\cite{BFG} and references therein).

In the limit of strong annihilation, $\mbox{Im }a_{s}$ depends neither 
on the range of interaction $R$ nor on its particular form within this 
region, but it is determined only by the diffuseness of the exponential 
tail $r_{0}$. The wave-function $\Psi$ exponentially decreases when $r$
crosses the region of diffuseness of the potential:
\begin{eqnarray*}
\Psi \sim \exp \left(\int_{0}^{r}\sqrt{2m|U(\rho )|}d\rho \right)
\end{eqnarray*}
From the expression for the scattering amplitude:
\begin{eqnarray*}
f(k) =-\frac{2m}{k^{2}}\int_{0}^{\infty} \sin \left( kr\right) U(r)\Psi (r)dr
\end{eqnarray*}
($k$ being the center of mass momentum) one can see that the 
contribution of the inner part of the interaction is exponentially small 
due to the behavior of the wave-function. Qualitatively this means that the 
particle in a strong imaginary potential ``annihilates'' at a large 
distance and that the probability to penetrate to the origin is 
exponentially small. The scattering takes place on the exponential tail 
and is thus determined by $r_{0}$. 

As was shown in \cite{FG99}, this phenomenon leads to the effect of 
saturation of the antiproton-nucleus and kaon-nucleus atomic widths.

Another prediction can be made for the value of the antinucleon-nucleus 
annihilation cross-sections. At very low energies, where the 
annihilation cross-section $\sigma_{a}$ is determined by $\mbox{Im 
}a_{s}$: $k\sigma_{a}\approx 4\pi \,|\mbox{Im }a_{s}|$, its value does 
not depend on the atomic number (for nuclei with the same 
diffuseness).  Strictly speaking, this simple conclusion is true for 
\={n} only (for \={p}, the low energy expression for the annihilation 
cross-section depends on the real part of the scattering length and 
contains higher partial wave parameters \cite{PBLZ,CPZ}, so that this case 
needs more careful consideration).

To complete this analysis let us consider an elementary interaction in a 
separable form $V = g \, |\alpha\rangle \langle \alpha |$ with complex 
coupling constant $g$ (as chosen in \cite{Wycech85,LT}).  The 
imaginary part of the amplitude as a function of $g$ has the same 
qualitative one-bump behavior as the imaginary square well 
potential:
\begin{eqnarray*}
a_{s}= -\frac{g|\alpha (\vec{k} =0)|^2}
{1 + 2mg \int d^3 \vec{k}|\alpha (\vec{k})|^2/k^2}.
\end{eqnarray*}
where $\alpha (\vec{k} )= \langle\vec{k}| \alpha \rangle$.

For finite complex $g$, $a_{s}$ is complex also. However, for a very 
strong interaction $|g| \rightarrow \infty$, the scattering length 
becomes real for any form-factor $\langle \vec{k}|\alpha \rangle $.  
The reason for such behavior is the ``poor'' spectral properties of 
separable interactions: a rank-one separable potential is able to produce 
only one bound state and all dynamics is governed by the position of 
the corresponding poles of the S-matrix. 

Till now, only potentials which have the same range of the real and of 
the imaginary part were considered. However, the 
optical potentials describing elementary anti\-proton-proton interaction 
are quite different. For instance, in the case of the popular 
Kohno-Weise potential \cite{KW1,KW2} which reproduces rather well all 
data on low energy nucleon-antinucleon interaction, the range of the 
real part of the interaction is larger than that of the imaginary part 
(see Figure~2 in \cite{KW2}). For these potentials, the existence of an 
external real part of the potential can significantly modify the 
situation.

As an example, let's consider the simplest potential composed of two 
square wells where the internal potential is purely imaginary and 
external one real: 
\begin{eqnarray*}
U(r) = \left\{ 
\begin{array}{cl}
-iW, & r < R_1 \\ 
- V, & R_1 < r < R_2 \\ 
0, & r > R_2
\end{array}
\right. .
\end{eqnarray*}
The scattering length for this potential is given by 
\begin{eqnarray*}
a_{s} = R_2- \frac{1}{p} \tan \left[p(R_2-R_1) - \arctan \left\{
\frac{p\tan \kappa R_1 }{\kappa}\right\} \right].
\end{eqnarray*}
Here $\kappa^2 = 2imW$ and $p^2 = 2mV$.

If the external part of the potential is strong enough to create 
nearthreshold bound or resonant states, one obtains:
\begin{eqnarray*}
-\mbox{Im }a_{s}\approx \frac{1}{\sqrt{2mW}}\cdot
\frac{1}{\cos^{2}p(R_{2}-R_{1})}
\end{eqnarray*}
These states manifest themselves as an additional enhancement factor 
$1/\cos ^{2}p(R_{2}-R_{1})$.

In the case of an exact resonance (the bound state has an energy exactly 
equal to zero $p(R_{2}-R_{1})=(2n+1)\frac{\pi }{2}$) instead of a 
decreasing function seen previously we obtain an increasing one when 
$W\rightarrow \infty$
\begin{eqnarray*}
-\mbox{Im }a_{s}\approx \frac{\sqrt{mW}}{p^{2}}.
\end{eqnarray*}
The presence of near threshold poles (bound states or resonances) can 
significantly change the behaviour of $\mbox{Im }a_{s}$.

%%%%%%%%%%%%%%%%%%%%%%%%%%%%%%%%%%%%%%%%%%%%%%%%%%%%%

\section{\={p}D system}

\label{sec:3}

In the case of \={p}D scattering, it is possible to obtain numerical 
solution of the Faddeev equations. In this approach, all three body effects  
are taken into account. We performed such a calculation with a simple 
model of the \={N}N interaction to study the effect of strong annihilation and 
find out whether it is possible to reproduce the experimentally observed 
tendency. All interactions in this three body system were chosen in 
local form to avoid problems related to particular properties of 
separable interaction.

We solve the Faddeev equations in the coordinate space. The method can 
be found in \cite{3body}. Here we mention only main points.

%%%%%%%%%%%%%%%%%%%%%%%%%%%%%%%%%%%%%%%%%%%%%%%%%%%%%%%%

\subsection{Description of the model}

The three body wave function in the Faddeev approach is represented as the 
sum of three components 
\begin{eqnarray*}
\Psi= \Psi_1(\vec{x}_1,\vec{y}_1) + \Psi_2(\vec{x}_2,\vec{y}_2)+ \Psi_3
(\vec{x}_3,\vec{y}_3).
\end{eqnarray*}
where $\vec{x}_i,\vec{y}_i$ ($i=1,2,3$) denote three sets of Jacobi
coordinates defined according to 
\begin{eqnarray*}
\vec{x}_i&=& \vec{r}_j- \vec{r}_k, \\
\vec{y}_i&=& \frac{2}{\sqrt{3}} \left(\frac{\vec{r}_j+ \vec{r}_k}{2}-
\vec{r}_i\right).
\end{eqnarray*}
Here $\vec{r}_i$ is nucleon (or antinucleon) position vector.

The three-body Schr\"odinger equation is then rewritten as the system 
of equations for the Faddeev components: 
\begin{eqnarray}
&&\left[ k^{2}+\Delta _{\vec{x}_{1}}+\Delta _{\vec{y}_{1}}-m
V_{1}(\vec{x}_{1})\right] \Psi _{1}(\vec{x}_{1},\vec{y}_{1})=  \nonumber
\label{eq1} \\
&&\hspace{1cm}mV_{1}(\vec{x}_{1})[\Psi _{2}
(\vec{x}_{2},\vec{y}_{2})+\Psi _{3}(\vec{x}_{3},\vec{y}_{3})],  \nonumber \\
&&\left[ k^{2}+\Delta _{\vec{x}_{2}}+\Delta _{\vec{y}_{2}}-m
V_{2}(\vec{x}_{2})\right] \Psi _{2}(\vec{x}_{2},\vec{y}_{2})=  \nonumber
\\
&&\hspace{1cm}mV_{2}(\vec{x}_{2})[\Psi _{3}
(\vec{x}_{3},\vec{y}_{3})+\Psi _{1}(\vec{x}_{1},\vec{y}_{1})],  \nonumber \\
&&\left[ k^{2}+\Delta _{\vec{x}_{3}}+\Delta _{\vec{y}_{3}}-m
V_{3}(\vec{x}_{3})\right] \Psi _{3}(\vec{x}_{3},\vec{y}_{3})=  \nonumber
\\
&&\hspace{1cm}mV_{3}(\vec{x}_{3})[\Psi _{1}
(\vec{x}_{1},\vec{y}_{1})+\Psi _{2}(\vec{x}_{2},\vec{y}_{2})].
\end{eqnarray}
where $E$ is the 3-body energy (above the mass of the $\bar{N}np$ 
system) and $k^{2}=mE$. To simplify the treatment, we choose the \={N}N
interaction in the
isospin independent form. Therefore we put $V_{1}(x)=V_{2}(x)=U(x)$ where 
$U$ is the complex antinucleon-nucleon potential. In this case the 
components $\Psi _{1} $ and $\Psi _{2}$ have the same functional 
dependence on their arguments. We put also $V_{3}(x)=V_{\mbox{\tiny 
d}}(x)$ where $V_{\mbox{\tiny d}}(x)$ is the $np $ potential responsible for 
describing the deuteron. For $V_{\mbox{\tiny d}}(x)$ we use the 
Mafliet-Tjon potential \cite{MT}. It corresponds to the S-wave $np$ 
interaction only. Since we consider the low energy scattering, we keep 
also only the S-wave in the antinucleon-deuteron scattering. Hence, the 
Faddeev components depend on the moduli $x_{i},y_{i}$.  Instead of 
$\Psi _{i}(x_{i},y_{i})$ we introduce $\psi _{i}(x_{i},y_{i})$: 
\begin{eqnarray*}
\Psi _{i}(x_{i},y_{i})=\frac{\displaystyle{\psi _{i}(x_{i},
y_{i})}}{\displaystyle{x_{i}y_{i}}}. 
\end{eqnarray*}
In the r.h.-sides of the equations (\ref{eq1}), the moduli of the
Jacobi coordinates $x_{2},y_{2}$ and $x_{3},y_{3}$ expressed in
terms of $\vec{x}_{1},\vec{y}_{1}$, depend on the angle $\theta $ between
$\vec{x}_{1}$ and $\vec{y}_{1}$. Therefore the partial wave decomposition
results in the integrals over $u=\cos \theta $. In this
way, the equations (\ref{eq1}) turn into the two component system of
integro-differential equations: 
\begin{eqnarray}
&&\left[ k^{2}+\frac{\partial ^{2}}{\partial x^{2}}+\frac{\partial ^{2}}
{\partial y^{2}}-mU(x)\right] \psi _{1}(x,y)=  \nonumber
\label{eq2} \\
&&\hspace{1cm}mU(x)\frac{1}{2}\int_{-1}^{+1}du\frac{xy}
{x^{\prime }y^{\prime }}\left[ \psi _{1}(x^{\prime },y^{\prime })+\psi
_{3}(x^{\prime },y^{\prime })\right]  \nonumber \\
&&\left[ k^{2}+\frac{\partial ^{2}}{\partial x^{2}}+\frac{\partial ^{2}}
{\partial y^{2}}-mV_{\mbox{\tiny d}}(x)\right] \psi
_{3}(x,y)=  \nonumber \\
&&\hspace{1cm}mV_{\mbox{\tiny d}}(x)\int_{-1}^{+1}du
\frac{xy}{x^{\prime }y^{\prime }}\psi _{1}(x^{\prime },y^{\prime }),
\end{eqnarray}
with 
\begin{eqnarray*}
x^{\prime } &=&\frac{1}{2}(x^{2}+2\sqrt{3}xyu+3y^{2})^{1/2}, \\
y^{\prime } &=&\frac{1}{2}(3x^{2}-2\sqrt{3}xyu+y^{2})^{1/2}.
\end{eqnarray*}
We have taken into account that the dependence of the Faddeev 
components in the integrands on two sets of variables (e.g., on 
$x_{2},y_{2}$ and $x_{3},y_{3}$ in the first equation in (\ref{eq1})) 
is reduced, because of integration over $u$, to the single set
$x^{\prime },y^{\prime }$.

Below the three-body threshold, in the asymptotic region of $x,y$ the 
component $\psi _{1}(x,y)$ decreases exponentially. The component $\psi 
_{3}(x_{,}y)$ describes the system asymptotically
(when the distance $\sqrt{3}y/2$ between antiproton and deu\-teron is large): 
\begin{equation}
\psi _{1}(x,y\rightarrow \infty )=\varphi _{\mbox{\tiny d}}(x)\left[ \sin
(yq)+f(q)\exp (iqy)\right] .
\end{equation}
Here $\varphi _{\mbox{\tiny d}}(x)$ is the deuteron wave function, 
$f(q)$ is the antiproton-deuteron scattering amplitude which is related 
to the corresponding phase shift $\delta $ by $f(q)=(\exp (2i\delta 
)-1)/(2i)$.  The momentum $q$ is related to the center-of-mass 
\={p}D energy as $E_{cm}^{\mbox{\small \={p}D}}=q^{2}/m$. With these 
definitions of $f(q)$ and $q$ the scattering length is found to be: 
\begin{eqnarray*}
a=-\frac{\displaystyle\sqrt{3}}{\displaystyle{2}}\lim_{q\rightarrow 0}
\frac{\displaystyle{f(q)}}{\displaystyle{q}}. 
\end{eqnarray*}
The factor $\sqrt{3}/2$ appears since it enters the relation between 
the Jacobi coordinate $\vec{y}$ and the relative \={p}D distance.

The $\bar{N}N$ potential was taken in the Woods-Saxon form to reproduce 
main features of Kohno-Wiese optical potential \cite{KW1,KW2}: 
\begin{eqnarray*}
U(r) = &&- V \left[1 + \exp \left\{\frac{r-R_r}{r_r}\right\} \right]^{-1} \\
&&-iW \left[1 + \exp \left\{\frac{r-R_i}{r_i}\right\} \right]^{-1}.
\end{eqnarray*}
with $R_r= 1.2$ fm, $r_r=$ 0.3 fm, $R_i= 0.55$ fm, $r_i=$ 0.2 fm (to 
reproduce the fact that the range of the real part is bigger than that 
of the imaginary part). $V$ and $W$ are considered as parameters. The 
imaginary part of the interaction is exactly the same as in the original 
potential \cite{KW1,KW2}.

%%%%%%%%%%%%%%%%%%%%%%%%%%%%%%%%%%%%%%%%%% 

\subsection{Pure annihilation and resonance limits}

\begin{figure}[h!]
\epsfxsize=8cm
\centerline{\epsfbox{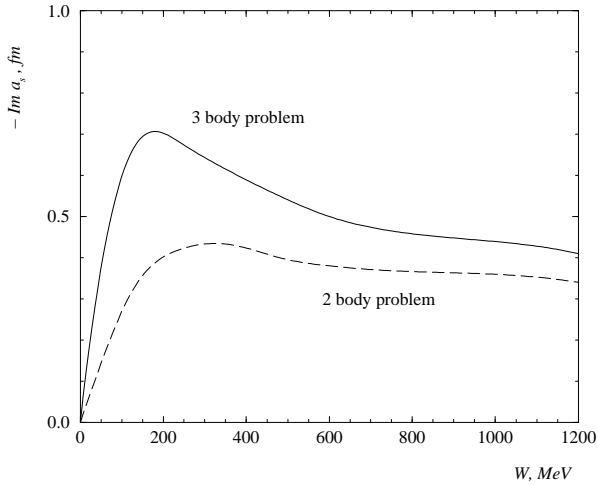}}
\caption{ The imaginary part of the scattering length as a function of 
the imaginary part $W$ of the Woods-Saxon potential when two body 
\={N}N interaction is pure annihilation one (two body problem - dashed 
line; three body problem - solid line) } \label{3annpur} \end{figure}

The simplest and very instructive case of the \={N}N interaction is the 
limit of pure annihilation $V=0$. The imaginary part of the scattering 
length as a function of the imaginary depth $W$ in two (\={N}N) and 
three (\={N}D) body problems are presented in Figure~\ref{3annpur}.

%-------------------------------------------------------------------------------
%
%-------------------------------------------------------------------------------

The results coincide with naive geometrical estimations: the imaginary 
part of the scattering length appears to be higher in the three body 
problem. For strong annihilation, the difference between the two values 
becomes smaller.  This result is quite natural when we take into 
account the arguments given in the Section 2: for large values of $W$, 
$\mbox{Im } a_{s}$ does not depend on the size of nucleus $R$ and is 
determined only by diffuseness.

This limit shows that the observed relation between $\mbox{Im }a_{s}$ 
for \={p}p and \={p}D interactions cannot be explained within a model 
of pure annihilation.

\begin{figure}[h!]
\epsfxsize=8cm
\centerline{\epsfbox{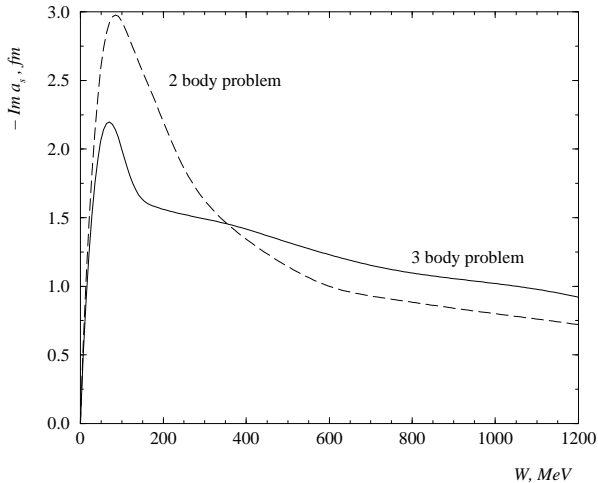}}
\caption{ The imaginary part of the scattering length as a function of 
the imaginary part $W$ of the Woods-Saxon potential when two body 
\={N}N resonances are close to the threshold }
\label{3res}
\end{figure}

To make $\mbox{Im }a_{s}$ for the \={p}p system bigger than for the 
\={p}D system, it is necessary to create a near threshold resonance or 
bound state (a pole of the S-matrix) in the two body system. 
Figure~\ref{3res} shows an example when the real part of the two body 
interaction $V$ is strong enough (80 MeV) to be able to produce a 
S-wave pole which is very close to the threshold.  Without annihilation 
($W=0$), one has a loosely bound state with the S-matrix pole position 
in $k$-plane of $k_b=i31.7$ MeV/$c$.  When annihilation is switched on 
this state acquires a width and is shifted away from the threshold: for 
$W = 50$ MeV, $k_b=-17.3+ i32.2$ MeV/$c$ and for $W = 500$ MeV, 
$k_b=-130.9+i72.9$ MeV/$c$.

This state manifests itself by quite a large value of $\mbox{Im } a_{s}$, 
as we discussed previously, in both two and three body problems. 
However, this resonance is less pronounced in the three body case and 
thus $\mbox{Im } a_{s}$ in this case is less sensitive to variation of 
the annihilation strength.

%-------------------------------------------------------------------------------
%
%-------------------------------------------------------------------------------

This example shows how one can change the relation between $\mbox{Im } 
a_{s}$ in two and three body problems.

\begin{figure}[h!]
\epsfxsize=8cm
\centerline{\epsfbox{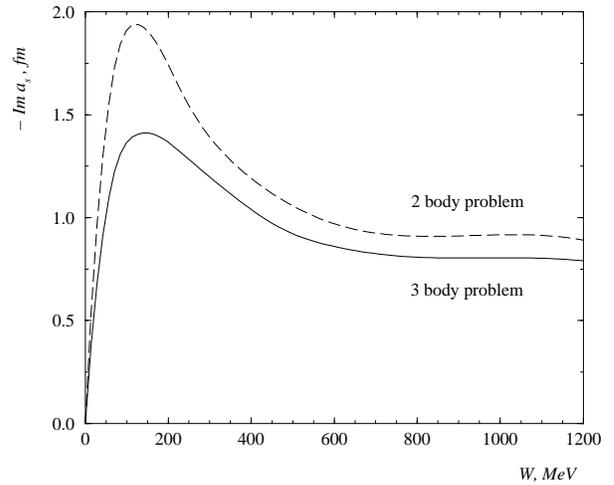}}
\caption{ The imaginary part of the scattering length as a function of 
the imaginary part $W$ of the Woods-Saxon potential when two body 
\={N}N resonances are far from the threshold }
\label{3farres}
\end{figure}

There are, at least, two ways to obtain a more realistic value of the 
imaginary part of the scattering length. Firstly, one can shift this 
state quite far from the threshold. Secondly, one can create not ground 
but an excited state which is sufficiently less sensitive to annihilation 
because of the zero in the wave function.  Figure~\ref{3farres} shows an 
example of such a situation ($V= 500$ MeV).  Without annihilation, the 
potential creates a loosely bound near threshold state (with $k_b=i36.5$ 
MeV/$c$) which is, in fact, the first excited state (the ground state 
is very deep $k_b=i485$ MeV/$c$). When the annihilation is switched on, 
the corresponding pole is shifted out of the threshold as previously 
but it is less sensitive to annihilation: for $W = 50$ MeV, $k_b=-11+ 
i37.8$ MeV/$c$ and for $W = 500$ MeV, $k_b=-46+i80.3$ MeV/$c$.

%-------------------------------------------------------------------------------
%
%-------------------------------------------------------------------------------

It is interesting to note that in Kohno-Weise model \cite{KW1,KW2} such 
quite large resonances appear in all 4 (!) partial waves (with two 
different spins and isospins) \cite{CDPS}. Moreover, one of them is an 
excited state.

As was found in \cite{LT2} these states appear also in a separable 
potential which was used for the three body calculations \cite{LT}.

\section{Conclusions}

\label{sec:4} 

Two- and three-body systems with very strong annihilation have been 
investigated using the optical model potential to understand the 
behaviour of the imaginary part of the scattering length as a function 
of the annihilation strength.

In two-body systems, the imaginary part of the scattering length is 
sensitive mostly to the diffuseness of the potential.  This 
circumstance gives a quite strong experimental prediction:  the low 
energy antineu\-tron-nucleus annihilation cross-section should be 
approximately the same for all nuclei.

Due to Coulomb forces, the situation in antiproton-nucleus annihilation 
can be more complicated: higher partial waves and the 
real part of the scattering amplitude could modify this simple picture.  
However, the experiments on the future AD (antiproton decelerator) facility 
would be of vital importance to understand the antinucleon-nucleus 
interaction. 

For three body systems, we propose a quite simple mechanism which is able 
to explain the observation that the imaginary part of the 
antiproton-deuteron scattering length is smaller than the 
antiproton-proton one.  This mechanism is based on two points. Firstly, 
the imaginary part of the scattering length in both systems are 
determined by the diffuseness of the annihilation potential. Thus these 
values are close to each other. Secondly, the two-body \={N}N interaction 
produces S-wave poles  and enhances the values of the scattering 
length. This enhancement is more important in the \={p}p system than in 
the \={p}D one. These S-wave poles (resonances) are necessarily present 
in all models which describe the experimental data, however, their 
experimental observation is a delicate task because of their large 
widths.

\section*{Acknowledgments}

The authors are indebted to J.~Carbonell and C.~Gignoux for many 
elucidating discussions on solving the Faddeev equations. One of the 
authors (K.V.P.) would like to thank A.~Zenoni, E.~Lodi Rizzini, and 
A.~Bianconi for useful discussions.  Two of the authors (V.A.K.  and 
A.Yu.V.) are sincerely grateful for the warm hospitality of the theory 
group at the Institut des Sciences Nucl\'{e}aires, Universit\'e Joseph 
Fourier, in Grenoble, where this work was performed.

\end{document}